\documentclass[aps,prl,twocolumn,superscriptaddress]{revtex4-1}

\usepackage{amsmath}
\usepackage[pdftex]{graphicx}
\usepackage[thinspace]{SIunits}

\providecommand \BibitemShut  [1]{\csname bibitem#1\endcsname}%

\begin{document}


\title{Dissipative preparation of the exciton and biexciton in a self-assembled quantum dot on picosecond timescales}
\author{Per-Lennart Ardelt}
 \affiliation{Walter Schottky Institut and Physik-Department, Technische Universit\"at M\"unchen, Am Coulombwall 4, 85748 Garching, Germany \\}
\author{Lukas Hanschke}
 \affiliation{Walter Schottky Institut and Physik-Department, Technische Universit\"at M\"unchen, Am Coulombwall 4, 85748 Garching, Germany \\}
 \author{Kevin A. Fischer}
 \affiliation{E. L. Ginzton Laboratory, Stanford University, Stanford, California 94305, USA\\}
 \author{Kai M\"uller}
 \affiliation{Walter Schottky Institut and Physik-Department, Technische Universit\"at M\"unchen, Am Coulombwall 4, 85748 Garching, Germany \\}
 \affiliation{E. L. Ginzton Laboratory, Stanford University, Stanford, California 94305, USA\\}
\author{Alexander Kleinkauf}
 \affiliation{Walter Schottky Institut and Physik-Department, Technische Universit\"at M\"unchen, Am Coulombwall 4, 85748 Garching, Germany \\}
\author{Manuel Koller}
 \affiliation{Walter Schottky Institut and Physik-Department, Technische Universit\"at M\"unchen, Am Coulombwall 4, 85748 Garching, Germany \\}
\author{Alexander Bechtold}
 \affiliation{Walter Schottky Institut and Physik-Department, Technische Universit\"at M\"unchen, Am Coulombwall 4, 85748 Garching, Germany \\}
\author{Tobias Simmet}
 \affiliation{Walter Schottky Institut and Physik-Department, Technische Universit\"at M\"unchen, Am Coulombwall 4, 85748 Garching, Germany \\}
\author{Jakob Wierzbowski}
 \affiliation{Walter Schottky Institut and Physik-Department, Technische Universit\"at M\"unchen, Am Coulombwall 4, 85748 Garching, Germany \\}
\author{Hubert Riedl}
\affiliation{Walter Schottky Institut and Physik-Department, Technische Universit\"at M\"unchen, Am Coulombwall 4, 85748 Garching, Germany \\}
\author{Gerhard Abstreiter}
\affiliation{Institut for Adavanced Study, Technische Universit\"at M\"unchen, Lichtenbergstr. 2a, 85748 Garching, Germany \\}
\affiliation{Walter Schottky Institut and Physik-Department, Technische Universit\"at M\"unchen, Am Coulombwall 4, 85748 Garching, Germany \\}
\author{Jonathan. J. Finley}
 \affiliation{Walter Schottky Institut and Physik-Department, Technische Universit\"at M\"unchen, Am Coulombwall 4, 85748 Garching, Germany \\}
\email{finley@wsi.tum.de}

\date{\today}

\begin{abstract}
Pulsed resonant fluorescence is used to probe ultrafast phonon-assisted exciton and biexciton preparation in individual self-assembled InGaAs quantum dots. By driving the system using large area ($\geq10\pi$) near resonant optical pulses, we experimentally demonstrate how phonon mediated dissipation within the manifold of dressed excitonic states can be used to prepare the neutral exciton with a fidelity $\geq 70\%$. By comparing the phonon-assisted preparation with resonant Rabi oscillations we show that the phonon-mediated process provides the higher fidelity preparation for large pulse areas and is less sensitive to pulse area variations. Moreover, by detuning the laser with respect to the exciton transition we map out the spectral density for exciton coupling to the bulk LA-phonon continuum. Similar phonon mediated processes are shown to facilitate direct biexciton preparation via two photon biexciton absorption, with fidelities $>80\%$. Our results are found to be in very good quantitative agreement with simulations that model the quantum dot-phonon bath interactions with Bloch-Redfield theory.
\end{abstract}

\pacs{78.67.Hc 81.07.Ta 85.35.Be}

\maketitle

Due to their discrete electronic structure and strong interaction with light, self-assembled quantum dots (QDs) are often described as artificial atoms in the solid state. Indeed, many quantum optical experiments have recently been performed that exploit these atom-like properties; specific examples including deterministic single \cite{Michler2000} and entangled photon generation \cite{Michler2014}, Rabi oscillations \cite{Zrenner2002}, Mollow triplet physics \cite{Vamivakas2009}, optical spin pumping  \cite{atature2006}, coherent optical spin control \cite{Press2008, deGreve2011, Poem2011, Kodriano2012, muller2013} and, most recently, spin-photon entanglement  \cite{yamamoto2012, gao2012, steel2013}. Moreover, in cavity QED experiments\cite{reithmaier2004} remarkable effects such as photon-blockade or photon-tunneling have opened the way to exploit QDs to generate novel quantum states of light \cite{Faraon2008}. In all these experiments, the solid state environment of the QDs manifests itself primarily in coupling to acoustic phonons - an effect that is unwanted since it results in decoherence of the quantum state, particular examples include incoherent population transfer between a QD and a detuned microcavity mode \cite{Laucht2009} or intra-molecular tunneling in vertically stacked QD-molecules \cite{Nakaoka2006, Muller2012phonon}. Moreover, in coherent optical exciton control experiments, coupling to acoustic phonons dominates the damping of Rabi oscillations \cite{Forstner2003,Ramsay2010}, limiting state preparation and control fidelities and the scope for possible applications. However, in other cases the coupling to acoustic phonons was exploited instead, e.g. for the high-fidelity spin initialisation by tunnel ionisation \cite{ muller2012high} or for achieving population inversion in electrically driven quantum dots \cite{Petta2004}. Recently, Gl\"assl et al. \cite{glassl2013} proposed that high-fidelity preparation of excitonic states could be achieved by exploiting the coupling of the dressed excitonic states to a quasi continuum of vibrational modes. Their approach involves to combining the relative advantages of both Rabi oscillations and rapid adiabatic passage \cite{Wei2014,Mathew2014} by making use of phonon-mediated relaxation in the presence of a strong, near resonant pulsed optical field. 

In this paper we investigate the phonon-assisted preparation of neutral exciton ($X_{0}$) and biexciton (2X) states in QDs using pulsed resonant fluorescence and compare the results with coherent resonant excitation. We present resonant fluorescence measurements of the neutral exciton transition for ps-pulsed excitation and demonstrate Rabi oscillations for pulse areas up to $10 \, \pi$. When detuning the laser from the exciton transition, we observe phonon-assisted population inversion for positive detuning of $\sim 0-2 \, meV$ reaching a maximum state preparation of $\geq 70 \%$ for a detuning of $\sim0.7\,meV$. Via power and detuning dependent measurements, we probe the dynamics of this process and map out the spectrum of the exciton LA-phonon coupling. Similar methods are shown to facilitate the biexciton preparation with two-photon Rabi oscillations for resonant pulses and high fidelity phonon-assisted two-photon biexciton preparation for detuned pulses. Our results are in full accord with the findings of very recent theoretical \cite{ glassl2013} and experimental works \cite{Quilter2014,Bounouar2014} illustrating that dissipative processes can be used for reliable quantum state preparation.

\begin{figure}[t]
\includegraphics[width=1\columnwidth]{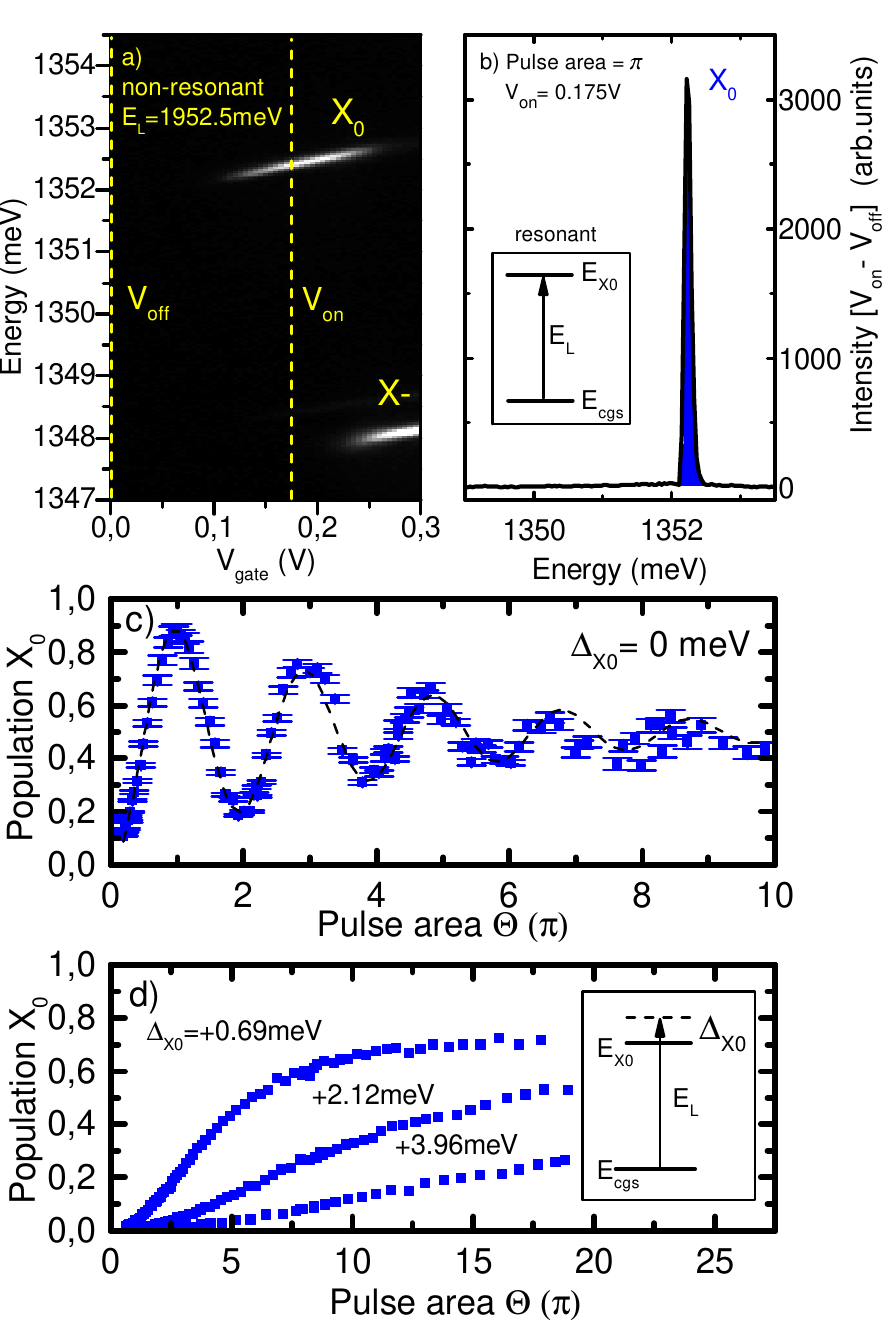}
\caption{\label{fig:Figure1} (Color online) (a) Voltage dependent PL measurement for above bandgap excitation at $1952.5 meV$. (b) Cross-polarised resonant fluorescence of the $X_{0}$ using pulsed excitation at $1352.2meV$. To eliminate a weak backround of the scattered laser, a reference spectrum at $V_{off}$ is substracted from the on-resonance spectrum at $V_{on}$ (c) Rabi oscillations for resonantly driving the exciton transition $X_{0}$ (illustrated in the inset in (b)) with pulses of a length of $\sim 10 ps$. (d) Phonon-assisted excitation of the exciton state for different laser detunings $\Delta_{X0}$ from the neutral excitation $X_{0}$.}
\end{figure}

The sample investigated is grown by molecular beam epitaxy (MBE). On top of a 14-pair $\lambda / 4$ GaAs/AlAs distributed Bragg reflector (DBR) a layer of low density InGaAs QDs is centered in a nominally $260 \, nm$ thick GaAs layer acting as a weak micro cavity resulting in a high photon extraction efficiency from the surface of the sample \cite{gao2012}. A $95 \, nm$ thick n-doped layer below the QDs and a Ti/Au metal top contact form a Schottky diode that facilitates control of the electric field and, thereby, the energies of excitonic transitions in the QD using the DC Stark shift \cite{warburton2000}. The separation between the QDs and the doped layer is chosen to be $35 \, nm$ to achieve charge stability and sequential charging with electrons due to tunneling between the QD and the electron reservoir. Typical voltage dependent photoluminescence (PL) measurements of a single QD using above bandgap excitation at $ E_{Laser}=1952.5 \, meV$ are presented in figure \ref{fig:Figure1}a.  The data shows clear charging plateaus for the neutral exciton transition $X_{0}$ and the negatively charged exciton $X^{-}$ with charge stability for the neutral exciton $X_{0}$ from $V_{gate}=0.12 \,V$ to $V_{gate}=0.24 \,V$. We note here that all data presented throughout this manuscript were obtained at a lattice temperature of $T=4.2 \, K$.

For the remainder of the manuscript we focus on resonant fluorescence (RF) spectroscopy with linearly cross-polarised suppression of the excitation laser as described in detail in references \cite{Kuhlmann2013,He2013}. Thereby, we used a fs-pulsed Ti:Sa laser and 4f pulse-shaping to obtain tunable picosecond pulses with a temporal bandwidth of $\sim 10 \, ps$ \cite{Ramsay2010, Zecherle2010, muller2013} corresponding to a spectral bandwidth of $\sim 0.22 \, meV$. The luminescence intensity obtained from the neutral exciton $X_{0}$ under resonant pulsed excitation is presented in figure \ref{fig:Figure1}b for a fixed bias of $V_{gate}=0.175\, V$. Note, that in order to fully eliminate light scattered from the laser, we record a spectrum at the voltage of interest $V_{on}=0.175\,V$ and then subtract a spectrum obtained at $V_{off}=0\, V$, where PL emission is fully quenched (see Fig.\ref{fig:Figure1}a).  In this way, the RF signal colour coded in blue in Fig.\ref{fig:Figure1}b stems only from the QD and its intensity is directly proportional to the neutral exciton population $X_{0}$. It is well known, that for a laser field at energy $E_{L}$ resonantly driving the neutral exciton transition $E_{X0} = E_{L}$ (Fig \ref{fig:Figure1}b - inset), the coherent interaction between the laser field and the QD will lead to population oscillations between the crystal ground state $cgs$ and the neutral exciton $X_{0}$ as well as dressing of the exciton state \cite{Forstner2003,Zrenner2002}. To investigate the coherent interaction between the laser and QD as well as the possibility of phonon-assisted population inversion, we present in figure \ref{fig:Figure1}c the population of the neutral exciton $X_{0}$ as a function of the pulse area $\theta$ of the driving laser field. To extract the exciton population $X_{0}$, we normalized the intensity of the RF signal $I_{RF}$ to the intensity of a $\pi$ pulse corrected for damping corresponding to an initial amplitude $I_{0}$ \cite{Zrenner2002}. The normalized intensity $I_{RF}/I_{0}$ corresponding to the $X_{0}$ population is plotted with a guide to eye (dashed line) in Fig. \ref{fig:Figure1}c, revealing clear, damped Rabi oscillations up to pulse areas of $10 \pi$ confirming the coherent nature of the light-matter interaction. Note that due to the damping we measure a maximum exciton population $X_{0}$ of $89 \%$ for resonant excitation upon exciting the system with a $\pi$ pulse. Significant damping of the Rabi oscillations in figure \ref{fig:Figure1}c is observed with increasing pulse area due to coupling to acoustic phonons \cite{Forstner2003, Ramsay2010} with the population decreasing towards the incoherent limit of $\sim 50 \%$ at the highest pulse areas explored.   

We continue to perform measurements for different energy detunings $\Delta_{X0} = E_{L} - E_{X0}$  between driving laser and the neutral exciton transition $X_{0}$ (Fig \ref{fig:Figure1}d - inset). The results are presented in figure \ref{fig:Figure1}d that shows the population of the neutral exciton state $X_{0}$ as a function of the pulse area for typical detunings of $ \Delta_{X0}=+ 0.69 \, meV$, $ +2.12\,meV$ and $+ 3.96 \, meV$ respectively. Note that all the detunings $\Delta_{X0}$ presented are significantly larger than the bandwidth of the driving laser pulse of $\sim 0.22\, meV $. In contrast to the situation on resonance, a continuously increasing and saturating population is observed with increasing pulse area $\theta$ rather than Rabi oscillations.  A maximum population of $72\%$ is reached for a detuning of $\Delta_{X0} = +0.69\,meV$, well above the inversion limit of $50\%$ and only $17\%$ smaller than the maximum population created via a resonant $\pi$-pulse (Fig \ref{fig:Figure1}c). We note that for larger $\Delta_{X0}$ the increase slows and saturates at a lower maximum population. This behaviour is well described by a recent theoretical proposal from Gl\"assl et al.\cite{glassl2013} whereby phonon-assisted dissipation enhances the achievable exciton population for positive $ \Delta_{X0}$. The energy separation of the dressed states is increased by the optical field into a range where efficient coupling to LA-phonons occurs \cite{Vagov2011,Quilter2014} facilitating dissipative relaxation into the energetically lower lying dressed state that has predominantly excitonic character.

\begin{figure}[t]
\includegraphics[width=1\columnwidth]{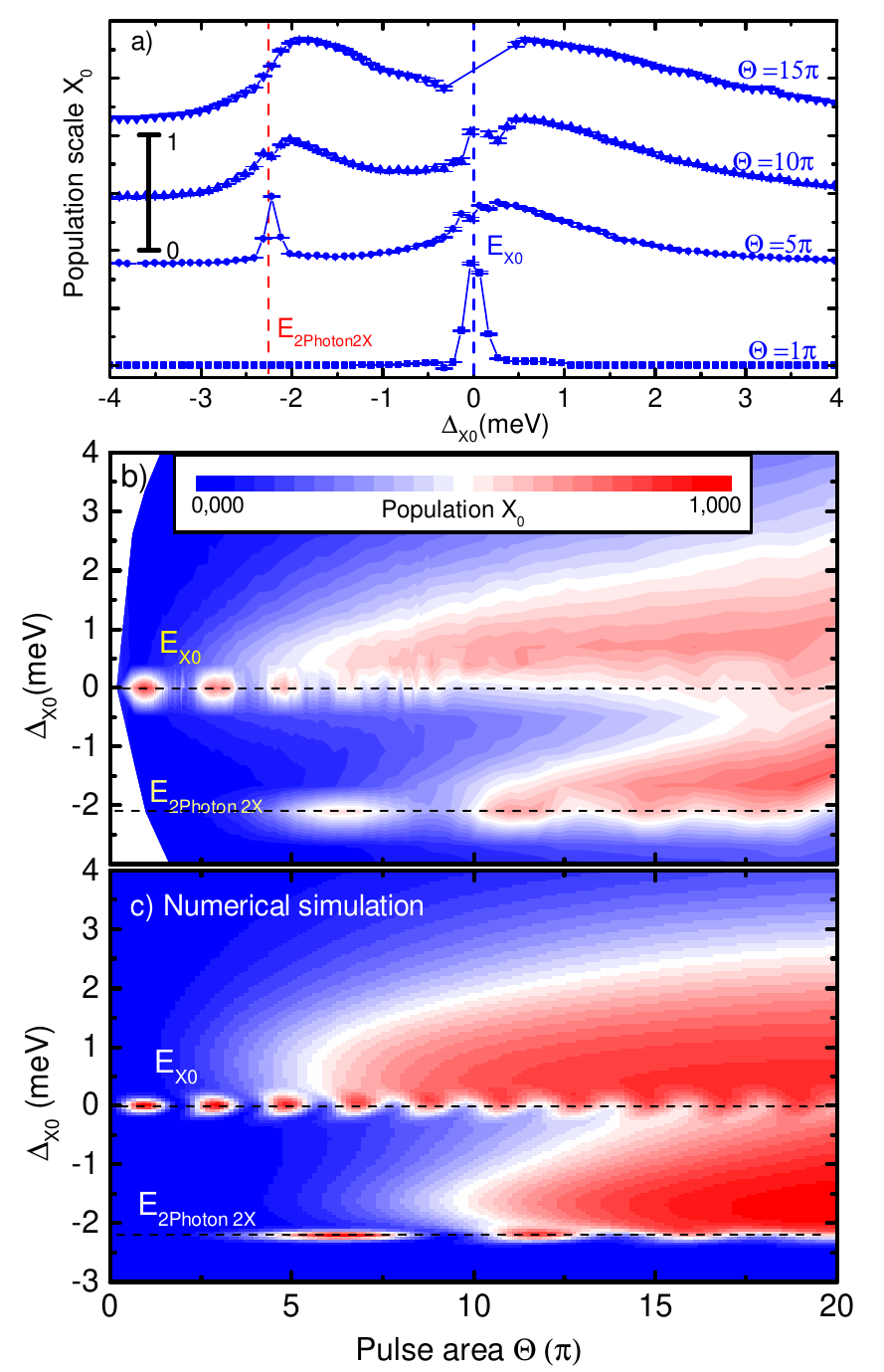}
\caption{\label{fig:Figure2} (Color online) (a) Population of the $X_{0}$ as a function of the laser detuning $\Delta_{X0}$ from the exciton $X_{0}$ for pulse areas of $1\pi$, $5\pi$, $10\pi$ and $15\pi$ (b) Countour plot of the neutral exciton population $X_{0}$ measured as function laser detuning and excitation pulse area and (c) theoretically calculated exciton population $X_{0}$ using a numerical simulation on the same colour scale.}
\end{figure}

To investigate the fidelity and energy range of phonon-assisted state preparation, we performed experiments with a fixed laser pulse area while varying the detuning between laser and exciton transition $\Delta_{X_{0}}=E_{L}-E_{X0}$. Selected results of these experiments are presented in figure \ref{fig:Figure2}a, showing the $X_{0}$ population as a function of the laser energy detuning $\Delta_{X0}$ for pulse areas of $1\pi$, $5\pi$, $10\pi$ and $15 \pi$, vertically offset for clarity. Figure \ref{fig:Figure2}b shows a false colour image of all data collected.  For a pulse area $\theta=1 \, \pi$ only a sharp resonance is observed at the energy of the exciton transition $E_{X0}$. In contrast, with increasing power this resonance broadens and becomes asymmetric due to the coupling to LA-phonons and low temperature that prohibits phonon absorption \cite{glassl2013}. The detuning, for which a maximum exciton population is generated, shifts to $\Delta_{X0}=+0.69 \, meV$ for pulse areas of $15\pi$, since for resonant excitation at $\Delta_{X0} = 0 meV$ the damping of the Rabi oscillations reduces the population to $\sim50 \%$, whereby phonon-assisted exciton generation becomes more efficient for $\Delta_{X0}\geq0$. We measure a maximum exciton population $X_{0}$ via phonon-assisted preparation of $ P_{(\Delta_{X0}=+0.32meV)} = 53 \pm 0.8 \%$, $P_{(\Delta_{X0}=+0.56meV)} = 68 \pm 1.1 \%$ and $P_{(\Delta_{X0}=+0.69meV)} = 0.72 \pm 0.8 \%$ for $5\pi$, $10\pi$ and $15\pi$ pulses, respectively. Since these values also directly correspond to the state preparation fidelity, we note that for larger pulse areas than $10\pi$, the phonon-assisted route guarantees higher state preparation fidelities than resonant excitation. 

\begin{figure}[t]
\includegraphics[width=1\columnwidth]{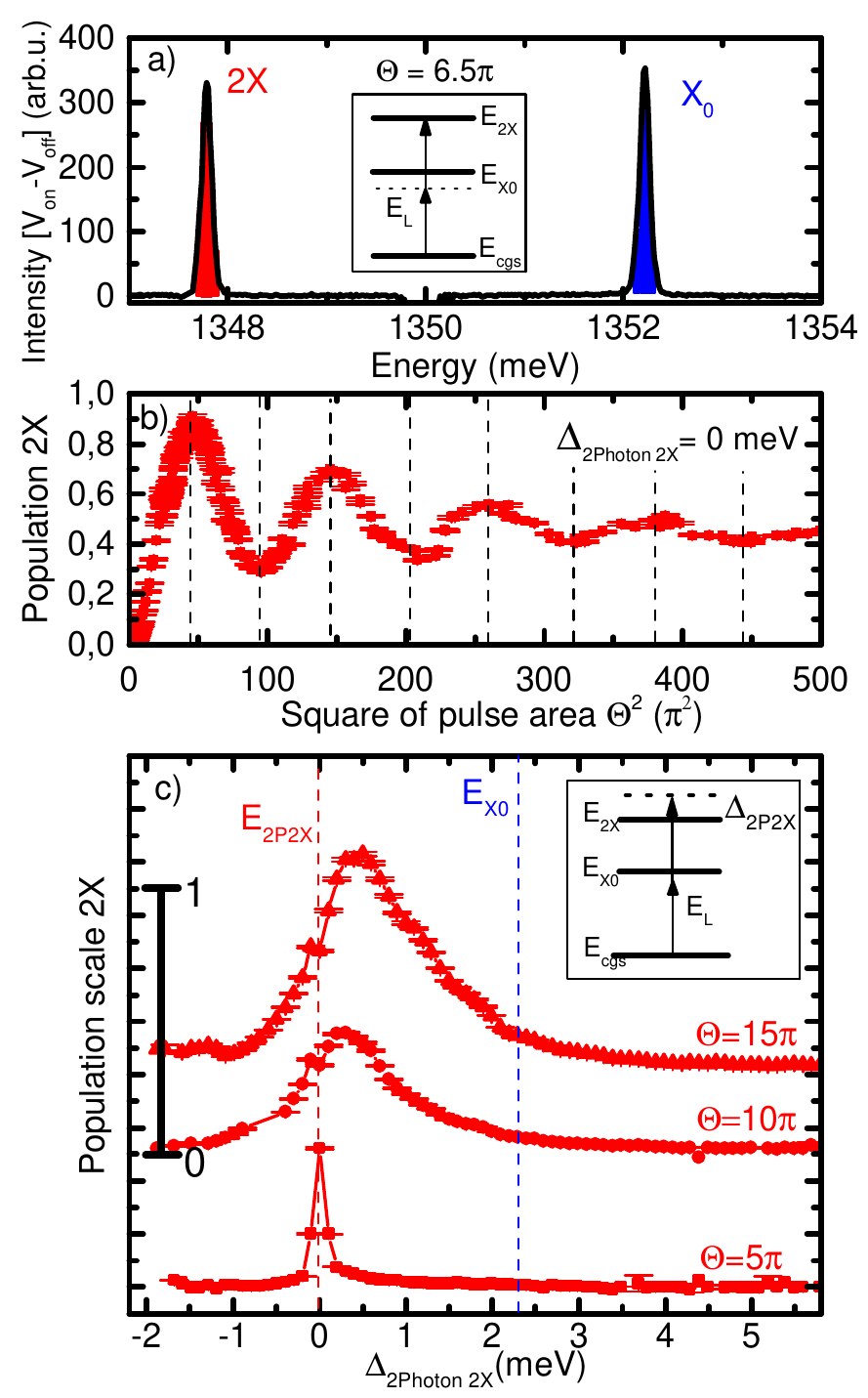}
\caption{\label{fig:Figure3} (Color online) (a) Energy spectrum for resontly driving the 2 photon 2X transition (illustrated in the inset) showing luminescence from the biexciton $2X$ (red) and exciton $X_{0}$ (blue) transitions.(b) Rabi-type oscillations of the biexciton 2X population resonantly driving the 2 photon 2X transition as a function of the square of pulse area from the Rabi oscillation presented in figure \ref{fig:Figure1}c. (c) Phonon assisted creation of the biexcition 2X population as a function of detuning (illustrated in the inset) from the 2 photon 2X resonance $\Delta_{2P2X}$ }\end{figure}

For pulse areas of $\theta = 5\pi$, $10\pi$ and $15\pi$, a second resonance is observed at a detuning of $\Delta_{X0}=-2.2 \, meV$, labelled $E_{2Photon2X}$ in Figs. \ref{fig:Figure2}a and \ref{fig:Figure2}b. Similar to the neutral exciton resonance, this feature exhibits a sharp peak for a pulse area $\theta = 5 \pi$ and asymmetrically broadens for larger pulse areas. Energetically $E_{2Photon2X}$ lies exactly midway between $X_{0}$ and $2X$, which are seperated by $\Delta_{bind}=- 4.4 \, meV$ (see Fig. \ref{fig:Figure3}a) for the QD under investigation. Therefore, this feature is identified as arising from resonant two-photon excitation of the biexciton $2X$ as will be discussed further below. The neutral exciton is in this case populated by incoherent radiative decay from the biexciton $2X$. 

In order to support the identification of the experimental features observed in figs \ref{fig:Figure2}a and \ref{fig:Figure2}b we simulated the expected form of the detuning dependent RF-data.  Hereby, we first used the quantum toolbox in python (QuTiP \cite{qutip}) to diagonalize the free Hamiltonian including the time dependent driving of the laser pulse at each time step, including exciton and biexciton states, without coupling to phonons.  Thereafter, we computed the Bloch-Redfield tensor for each step \cite{cohen1992} representing the interaction with the phonon bath using a bulk phonon spectrum described in ref \cite{glassl2011}. The quantum state of the system was then evolved in the interaction picture to obtain the final state population presented in fig \ref{fig:Figure2}c. Excellent qualitative agreement is observed between the results of our simulations (fig \ref{fig:Figure2}c) and experiments (fig \ref{fig:Figure2}b) for a QD with electron and hole confinement lengths of $a_{e}=4.9\,nm$ and $a_{h}=2.1\,nm$. To further support our interpretation of the additional absorption line being due to the two-photon excitation of the biexciton, Fig. \ref{fig:Figure3}a shows the RF spectrum obtained when the laser excitation energy is set to $E_{L}=E_{2photon2X}=1350.0\,meV$ fulfilling the condition $E_{L} = (E_{2X}+E_{X0})/2$ as illustrated in the inset. The spectrum displays two sharp peaks, one at the energy of the $X_{0} \rightarrow (cgs)$ transition (blue shaded) and one at the energy of $2X \rightarrow X_{0}$ transition (red shaded), separated by a binding energy of $\Delta_{bind}=-4.4 \, meV$. To investigate the coherence of the 2 photon-biexciton $2X$ transition, we plot the integrated intensity of the biexciton $2X$ marked in red in Fig.\ref{fig:Figure3}a as a function of the square of the excitation pulse area for pulses resonant with the two-photon transition $\Delta_{2photon2X}=0\,meV$. As can be seen in figure \ref{fig:Figure3}b, well resolved damped Rabi-type oscillations are observed demonstrating coherent non-linear light-matter interaction \cite{Stufler2006}. Note that to extact the biexciton population $2X$, we normalize the intensity $I_{2X}$  to the intensity of a $\pi$ pulse corrected for damping $I_{0}$ and plot $I_{2X}/I_{0}$. Strikingly, the oscillations presented in Fig.\ref{fig:Figure3}b are not periodic with the pulse area $\theta$ like Rabi oscillations of the exciton transition $X_{0}$ but rather exhibit maxima that scale proportional to the square of it $\theta^{2}$. In adidition, the frequency shows a clear renormalization for high excitation powers of $\pi^{2} > 200$. Both effects support the identification of this feature as arising from a two photon process \cite{Stufler2006}.

Finally, we turn our attention to the phonon-assisted preparation of the biexciton $2X$. Figure \ref{fig:Figure3}c shows the measured $2X$ population while exciting with laser pulses detuned from the two-photon resonance by $\Delta_{2Photon2X}$ as illustrated in the inset. The traces in Fig.\ref{fig:Figure3}c were recorded using excitation powers corresponding to pulse areas of $5\pi$, $10\pi$ and 15$\pi$ similar to the spectra presented earlier for the exciton transition $X_{0}$ in Fig.\ref{fig:Figure2}a. While (phonon-assisted) excitation of the exciton transition $X_{0}$ results in PL from the exciton only, the (phonon-assisted) excitation of the two-photon biexciton transition results in PL from both $X_{0}$ and $2X$. For higher pulse areas of $5 \pi$, $10 \pi$ and $15 \pi$, the biexciton population $2X$ created by phonon-assisted two-photon absorption (Fig.\ref{fig:Figure3}c) has a similar form as the phonon-assisted exciton preparation population in Fig.\ref{fig:Figure2}a. In particular, for an excitation power of $15 \pi$ the intensity observed from the exciton transition is identical for phonon-assisted preparation via both exciton and biexciton states.  This further supports that the $X_{0}$ population observed in Fig.\ref{fig:Figure2}a for negative detunings $\Delta_{X0}$ arises from the decay from the biexciton to the neutral exciton $2X \rightarrow X_{0}$ after a phonon assisted two-photon biexciton $2X$ creation. We also note that the preparation of the biexciton $2X$ is possible with a high-fidelity above $80 \%$ as seen in Fig.\ref{fig:Figure3}c for an excitation pulse area of $15 \pi$ in a phonon-assisted process. 
The deterministic and fast high-fidelity preparation of a biexciton could be used for the generation of polarisation entangled photon pairs from the biexciton-exciton cascade \cite{Michler2014}. However, in contrast to a coherent excitation of the two photon resonance with a $\pi$ pulse, the phonon assisted state preparation does not require a high-precision control of the excitation power and, thereby, is less sensitive to decoherence, making this process promising for future quantum-optical applications.

In summary, we have presented ultrafast resonant fluorescence studies illustrating how dissipative processes can be used for high fidelity state preparation of excitons and biexcitons in a single quantum dot. For pulses resonant with the exciton transition or two photon biexciton transition we observed Rabi oscillations up to pulse areas of $10 \, \pi$. For detuned pulses, we demonstrated a high fidelity phonon-assisted excitation of exciton or biexciton states. Using detuning and power dependent measurements we demonstrated that phonon-assisted state preparation is possible with fidelities identical to resonant coherent excitation and mapped out the spectrum of the exciton to longitudinal acoustic phonon coupling. 

We gratefully acknowledge financial support from the DFG via SFB-631, Nanosystems Initiative Munich, the EU via S3 Nano and BaCaTeC. KM acknowledges financial support from the Alexander von Humboldt foundation and the ARO (grant W911NF-13-1-0309). KF acknowledges financial support from the Lu Stanford Graduate Fellowship and the United States Department of Defense NDSEG fellowship.

\bibliography{Papers}

\begin{thebibliography}{36}%
\makeatletter
\providecommand \@ifxundefined [1]{%
 \@ifx{#1\undefined}
}%
\providecommand \@ifnum [1]{%
 \ifnum #1\expandafter \@firstoftwo
 \else \expandafter \@secondoftwo
 \fi
}%
\providecommand \@ifx [1]{%
 \ifx #1\expandafter \@firstoftwo
 \else \expandafter \@secondoftwo
 \fi
}%
\providecommand \natexlab [1]{#1}%
\providecommand \enquote  [1]{``#1''}%
\providecommand \bibnamefont  [1]{#1}%
\providecommand \bibfnamefont [1]{#1}%
\providecommand \citenamefont [1]{#1}%
\providecommand \href@noop [0]{\@secondoftwo}%
\providecommand \href [0]{\begingroup \@sanitize@url \@href}%
\providecommand \@href[1]{\@@startlink{#1}\@@href}%
\providecommand \@@href[1]{\endgroup#1\@@endlink}%
\providecommand \@sanitize@url [0]{\catcode `\\12\catcode `\$12\catcode
  `\&12\catcode `\#12\catcode `\^12\catcode `\_12\catcode `\%12\relax}%
\providecommand \@@startlink[1]{}%
\providecommand \@@endlink[0]{}%
\providecommand \url  [0]{\begingroup\@sanitize@url \@url }%
\providecommand \@url [1]{\endgroup\@href {#1}{\urlprefix }}%
\providecommand \urlprefix  [0]{URL }%
\providecommand \Eprint [0]{\href }%
\providecommand \doibase [0]{http://dx.doi.org/}%
\providecommand \selectlanguage [0]{\@gobble}%
\providecommand \bibinfo  [0]{\@secondoftwo}%
\providecommand \bibfield  [0]{\@secondoftwo}%
\providecommand \translation [1]{[#1]}%
\providecommand \BibitemOpen [0]{}%
\providecommand \bibitemStop [0]{}%
\providecommand \bibitemNoStop [0]{.\EOS\space}%
\providecommand \EOS [0]{\spacefactor3000\relax}%
\providecommand \BibitemShut  [1]{\csname bibitem#1\endcsname}%
\let\auto@bib@innerbib\@empty
\bibitem [{\citenamefont {Michler}\ \emph {et~al.}(2000)\citenamefont
  {Michler}, \citenamefont {Kiraz}, \citenamefont {Becher}, \citenamefont
  {Schoenfeld}, \citenamefont {Petroff}, \citenamefont {Zhang}, \citenamefont
  {Hu},\ and\ \citenamefont {Imamoglu}}]{Michler2000}%
  \BibitemOpen
  \bibfield  {author} {\bibinfo {author} {\bibfnamefont {P.}~\bibnamefont
  {Michler}}, \bibinfo {author} {\bibfnamefont {A.}~\bibnamefont {Kiraz}},
  \bibinfo {author} {\bibfnamefont {C.}~\bibnamefont {Becher}}, \bibinfo
  {author} {\bibfnamefont {W.}~\bibnamefont {Schoenfeld}}, \bibinfo {author}
  {\bibfnamefont {P.}~\bibnamefont {Petroff}}, \bibinfo {author} {\bibfnamefont
  {L.}~\bibnamefont {Zhang}}, \bibinfo {author} {\bibfnamefont
  {E.}~\bibnamefont {Hu}}, \ and\ \bibinfo {author} {\bibfnamefont
  {A.}~\bibnamefont {Imamoglu}},\ }\href@noop {} {\bibfield  {journal}
  {\bibinfo  {journal} {Science}\ }\textbf {\bibinfo {volume} {290}},\ \bibinfo
  {pages} {2282} (\bibinfo {year} {2000})}\BibitemShut {NoStop}%
\bibitem [{\citenamefont {M{\"u}ller}\ \emph {et~al.}(2014)\citenamefont
  {M{\"u}ller}, \citenamefont {Bounouar}, \citenamefont {J{\"o}ns},
  \citenamefont {Gl{\"a}ssl},\ and\ \citenamefont {Michler}}]{Michler2014}%
  \BibitemOpen
  \bibfield  {author} {\bibinfo {author} {\bibfnamefont {M.}~\bibnamefont
  {M{\"u}ller}}, \bibinfo {author} {\bibfnamefont {S.}~\bibnamefont
  {Bounouar}}, \bibinfo {author} {\bibfnamefont {K.~D.}\ \bibnamefont
  {J{\"o}ns}}, \bibinfo {author} {\bibfnamefont {M.}~\bibnamefont
  {Gl{\"a}ssl}}, \ and\ \bibinfo {author} {\bibfnamefont {P.}~\bibnamefont
  {Michler}},\ }\href@noop {} {\bibfield  {journal} {\bibinfo  {journal}
  {Nature Photonics}\ }\textbf {\bibinfo {volume} {8}},\ \bibinfo {pages} {224}
  (\bibinfo {year} {2014})}\BibitemShut {NoStop}%
\bibitem [{\citenamefont {Zrenner}\ \emph {et~al.}(2002)\citenamefont
  {Zrenner}, \citenamefont {Beham}, \citenamefont {Stufler}, \citenamefont
  {Findeis}, \citenamefont {Bichler},\ and\ \citenamefont
  {Abstreiter}}]{Zrenner2002}%
  \BibitemOpen
  \bibfield  {author} {\bibinfo {author} {\bibfnamefont {A.}~\bibnamefont
  {Zrenner}}, \bibinfo {author} {\bibfnamefont {E.}~\bibnamefont {Beham}},
  \bibinfo {author} {\bibfnamefont {S.}~\bibnamefont {Stufler}}, \bibinfo
  {author} {\bibfnamefont {F.}~\bibnamefont {Findeis}}, \bibinfo {author}
  {\bibfnamefont {M.}~\bibnamefont {Bichler}}, \ and\ \bibinfo {author}
  {\bibfnamefont {G.}~\bibnamefont {Abstreiter}},\ }\href@noop {} {\bibfield
  {journal} {\bibinfo  {journal} {Nature}\ }\textbf {\bibinfo {volume} {418}},\
  \bibinfo {pages} {612} (\bibinfo {year} {2002})}\BibitemShut {NoStop}%
\bibitem [{\citenamefont {Vamivakas}\ \emph {et~al.}(2009)\citenamefont
  {Vamivakas}, \citenamefont {Zhao}, \citenamefont {Lu},\ and\ \citenamefont
  {Atat{\"u}re}}]{Vamivakas2009}%
  \BibitemOpen
  \bibfield  {author} {\bibinfo {author} {\bibfnamefont {A.~N.}\ \bibnamefont
  {Vamivakas}}, \bibinfo {author} {\bibfnamefont {Y.}~\bibnamefont {Zhao}},
  \bibinfo {author} {\bibfnamefont {C.-Y.}\ \bibnamefont {Lu}}, \ and\ \bibinfo
  {author} {\bibfnamefont {M.}~\bibnamefont {Atat{\"u}re}},\ }\href@noop {}
  {\bibfield  {journal} {\bibinfo  {journal} {Nature Physics}\ }\textbf
  {\bibinfo {volume} {5}},\ \bibinfo {pages} {198} (\bibinfo {year}
  {2009})}\BibitemShut {NoStop}%
\bibitem [{\citenamefont {Atat{\"u}re}\ \emph {et~al.}(2006)\citenamefont
  {Atat{\"u}re}, \citenamefont {Dreiser}, \citenamefont {Badolato},
  \citenamefont {H{\"o}gele}, \citenamefont {Karrai},\ and\ \citenamefont
  {Imamoglu}}]{atature2006}%
  \BibitemOpen
  \bibfield  {author} {\bibinfo {author} {\bibfnamefont {M.}~\bibnamefont
  {Atat{\"u}re}}, \bibinfo {author} {\bibfnamefont {J.}~\bibnamefont
  {Dreiser}}, \bibinfo {author} {\bibfnamefont {A.}~\bibnamefont {Badolato}},
  \bibinfo {author} {\bibfnamefont {A.}~\bibnamefont {H{\"o}gele}}, \bibinfo
  {author} {\bibfnamefont {K.}~\bibnamefont {Karrai}}, \ and\ \bibinfo {author}
  {\bibfnamefont {A.}~\bibnamefont {Imamoglu}},\ }\href@noop {} {\bibfield
  {journal} {\bibinfo  {journal} {Science}\ }\textbf {\bibinfo {volume}
  {312}},\ \bibinfo {pages} {551} (\bibinfo {year} {2006})}\BibitemShut
  {NoStop}%
\bibitem [{\citenamefont {Press}\ \emph {et~al.}(2008)\citenamefont {Press},
  \citenamefont {Ladd}, \citenamefont {Zhang},\ and\ \citenamefont
  {Yamamoto}}]{Press2008}%
  \BibitemOpen
  \bibfield  {author} {\bibinfo {author} {\bibfnamefont {D.}~\bibnamefont
  {Press}}, \bibinfo {author} {\bibfnamefont {T.~D.}\ \bibnamefont {Ladd}},
  \bibinfo {author} {\bibfnamefont {B.}~\bibnamefont {Zhang}}, \ and\ \bibinfo
  {author} {\bibfnamefont {Y.}~\bibnamefont {Yamamoto}},\ }\href {\doibase
  10.1038/nature07530} {\bibfield  {journal} {\bibinfo  {journal} {Nature}\
  }\textbf {\bibinfo {volume} {456}},\ \bibinfo {pages} {218} (\bibinfo {year}
  {2008})}\BibitemShut {NoStop}%
\bibitem [{\citenamefont {De~Greve}\ \emph {et~al.}(2011)\citenamefont
  {De~Greve}, \citenamefont {McMahon}, \citenamefont {Press}, \citenamefont
  {Ladd}, \citenamefont {Bisping}, \citenamefont {Schneider}, \citenamefont
  {Kamp}, \citenamefont {Worschech}, \citenamefont {Hoefling}, \citenamefont
  {Forchel},\ and\ \citenamefont {Yamamoto}}]{deGreve2011}%
  \BibitemOpen
  \bibfield  {author} {\bibinfo {author} {\bibfnamefont {K.}~\bibnamefont
  {De~Greve}}, \bibinfo {author} {\bibfnamefont {P.~L.}\ \bibnamefont
  {McMahon}}, \bibinfo {author} {\bibfnamefont {D.}~\bibnamefont {Press}},
  \bibinfo {author} {\bibfnamefont {T.~D.}\ \bibnamefont {Ladd}}, \bibinfo
  {author} {\bibfnamefont {D.}~\bibnamefont {Bisping}}, \bibinfo {author}
  {\bibfnamefont {C.}~\bibnamefont {Schneider}}, \bibinfo {author}
  {\bibfnamefont {M.}~\bibnamefont {Kamp}}, \bibinfo {author} {\bibfnamefont
  {L.}~\bibnamefont {Worschech}}, \bibinfo {author} {\bibfnamefont
  {S.}~\bibnamefont {Hoefling}}, \bibinfo {author} {\bibfnamefont
  {A.}~\bibnamefont {Forchel}}, \ and\ \bibinfo {author} {\bibfnamefont
  {Y.}~\bibnamefont {Yamamoto}},\ }\href {\doibase 10.1038/NPHYS2078}
  {\bibfield  {journal} {\bibinfo  {journal} {Nature Physics}\ }\textbf
  {\bibinfo {volume} {7}},\ \bibinfo {pages} {872} (\bibinfo {year}
  {2011})}\BibitemShut {NoStop}%
\bibitem [{\citenamefont {Poem}\ \emph {et~al.}(2011)\citenamefont {Poem},
  \citenamefont {Kenneth}, \citenamefont {Kodriano}, \citenamefont {Benny},
  \citenamefont {Khatsevich}, \citenamefont {Avron},\ and\ \citenamefont
  {Gershoni}}]{Poem2011}%
  \BibitemOpen
  \bibfield  {author} {\bibinfo {author} {\bibfnamefont {E.}~\bibnamefont
  {Poem}}, \bibinfo {author} {\bibfnamefont {O.}~\bibnamefont {Kenneth}},
  \bibinfo {author} {\bibfnamefont {Y.}~\bibnamefont {Kodriano}}, \bibinfo
  {author} {\bibfnamefont {Y.}~\bibnamefont {Benny}}, \bibinfo {author}
  {\bibfnamefont {S.}~\bibnamefont {Khatsevich}}, \bibinfo {author}
  {\bibfnamefont {J.}~\bibnamefont {Avron}}, \ and\ \bibinfo {author}
  {\bibfnamefont {D.}~\bibnamefont {Gershoni}},\ }\href@noop {} {\bibfield
  {journal} {\bibinfo  {journal} {Physical review letters}\ }\textbf {\bibinfo
  {volume} {107}},\ \bibinfo {pages} {087401} (\bibinfo {year}
  {2011})}\BibitemShut {NoStop}%
\bibitem [{\citenamefont {Kodriano}\ \emph {et~al.}(2012)\citenamefont
  {Kodriano}, \citenamefont {Schwartz}, \citenamefont {Poem}, \citenamefont
  {Benny}, \citenamefont {Presman}, \citenamefont {Truong}, \citenamefont
  {Petroff},\ and\ \citenamefont {Gershoni}}]{Kodriano2012}%
  \BibitemOpen
  \bibfield  {author} {\bibinfo {author} {\bibfnamefont {Y.}~\bibnamefont
  {Kodriano}}, \bibinfo {author} {\bibfnamefont {I.}~\bibnamefont {Schwartz}},
  \bibinfo {author} {\bibfnamefont {E.}~\bibnamefont {Poem}}, \bibinfo {author}
  {\bibfnamefont {Y.}~\bibnamefont {Benny}}, \bibinfo {author} {\bibfnamefont
  {R.}~\bibnamefont {Presman}}, \bibinfo {author} {\bibfnamefont
  {T.}~\bibnamefont {Truong}}, \bibinfo {author} {\bibfnamefont
  {P.}~\bibnamefont {Petroff}}, \ and\ \bibinfo {author} {\bibfnamefont
  {D.}~\bibnamefont {Gershoni}},\ }\href@noop {} {\bibfield  {journal}
  {\bibinfo  {journal} {Physical Review B}\ }\textbf {\bibinfo {volume} {85}},\
  \bibinfo {pages} {241304} (\bibinfo {year} {2012})}\BibitemShut {NoStop}%
\bibitem [{\citenamefont {M{\"u}ller}\ \emph {et~al.}(2013)\citenamefont
  {M{\"u}ller}, \citenamefont {Kaldewey}, \citenamefont {Ripszam},
  \citenamefont {Wildmann}, \citenamefont {Bechtold}, \citenamefont {Bichler},
  \citenamefont {Koblm{\"u}ller}, \citenamefont {Abstreiter},\ and\
  \citenamefont {Finley}}]{muller2013}%
  \BibitemOpen
  \bibfield  {author} {\bibinfo {author} {\bibfnamefont {K.}~\bibnamefont
  {M{\"u}ller}}, \bibinfo {author} {\bibfnamefont {T.}~\bibnamefont
  {Kaldewey}}, \bibinfo {author} {\bibfnamefont {R.}~\bibnamefont {Ripszam}},
  \bibinfo {author} {\bibfnamefont {J.}~\bibnamefont {Wildmann}}, \bibinfo
  {author} {\bibfnamefont {A.}~\bibnamefont {Bechtold}}, \bibinfo {author}
  {\bibfnamefont {M.}~\bibnamefont {Bichler}}, \bibinfo {author} {\bibfnamefont
  {G.}~\bibnamefont {Koblm{\"u}ller}}, \bibinfo {author} {\bibfnamefont
  {G.}~\bibnamefont {Abstreiter}}, \ and\ \bibinfo {author} {\bibfnamefont
  {J.}~\bibnamefont {Finley}},\ }\href@noop {} {\bibfield  {journal} {\bibinfo
  {journal} {Scientific reports}\ }\textbf {\bibinfo {volume} {3}} (\bibinfo
  {year} {2013})}\BibitemShut {NoStop}%
\bibitem [{\citenamefont {De~Greve}\ \emph {et~al.}(2012)\citenamefont
  {De~Greve}, \citenamefont {Yu}, \citenamefont {McMahon}, \citenamefont
  {Pelc}, \citenamefont {Natarajan}, \citenamefont {Kim}, \citenamefont {Abe},
  \citenamefont {Maier}, \citenamefont {Schneider}, \citenamefont {Kamp} \emph
  {et~al.}}]{yamamoto2012}%
  \BibitemOpen
  \bibfield  {author} {\bibinfo {author} {\bibfnamefont {K.}~\bibnamefont
  {De~Greve}}, \bibinfo {author} {\bibfnamefont {L.}~\bibnamefont {Yu}},
  \bibinfo {author} {\bibfnamefont {P.~L.}\ \bibnamefont {McMahon}}, \bibinfo
  {author} {\bibfnamefont {J.~S.}\ \bibnamefont {Pelc}}, \bibinfo {author}
  {\bibfnamefont {C.~M.}\ \bibnamefont {Natarajan}}, \bibinfo {author}
  {\bibfnamefont {N.~Y.}\ \bibnamefont {Kim}}, \bibinfo {author} {\bibfnamefont
  {E.}~\bibnamefont {Abe}}, \bibinfo {author} {\bibfnamefont {S.}~\bibnamefont
  {Maier}}, \bibinfo {author} {\bibfnamefont {C.}~\bibnamefont {Schneider}},
  \bibinfo {author} {\bibfnamefont {M.}~\bibnamefont {Kamp}},  \emph {et~al.},\
  }\href@noop {} {\bibfield  {journal} {\bibinfo  {journal} {Nature}\ }\textbf
  {\bibinfo {volume} {491}},\ \bibinfo {pages} {421} (\bibinfo {year}
  {2012})}\BibitemShut {NoStop}%
\bibitem [{\citenamefont {Gao}\ \emph {et~al.}(2012)\citenamefont {Gao},
  \citenamefont {Fallahi}, \citenamefont {Togan}, \citenamefont
  {Miguel-Sanchez},\ and\ \citenamefont {Imamoglu}}]{gao2012}%
  \BibitemOpen
  \bibfield  {author} {\bibinfo {author} {\bibfnamefont {W.}~\bibnamefont
  {Gao}}, \bibinfo {author} {\bibfnamefont {P.}~\bibnamefont {Fallahi}},
  \bibinfo {author} {\bibfnamefont {E.}~\bibnamefont {Togan}}, \bibinfo
  {author} {\bibfnamefont {J.}~\bibnamefont {Miguel-Sanchez}}, \ and\ \bibinfo
  {author} {\bibfnamefont {A.}~\bibnamefont {Imamoglu}},\ }\href@noop {}
  {\bibfield  {journal} {\bibinfo  {journal} {Nature}\ }\textbf {\bibinfo
  {volume} {491}},\ \bibinfo {pages} {426} (\bibinfo {year}
  {2012})}\BibitemShut {NoStop}%
\bibitem [{\citenamefont {Schaibley}\ \emph {et~al.}(2013)\citenamefont
  {Schaibley}, \citenamefont {Burgers}, \citenamefont {McCracken},
  \citenamefont {Duan}, \citenamefont {Berman}, \citenamefont {Steel},
  \citenamefont {Bracker}, \citenamefont {Gammon},\ and\ \citenamefont
  {Sham}}]{steel2013}%
  \BibitemOpen
  \bibfield  {author} {\bibinfo {author} {\bibfnamefont {J.}~\bibnamefont
  {Schaibley}}, \bibinfo {author} {\bibfnamefont {A.}~\bibnamefont {Burgers}},
  \bibinfo {author} {\bibfnamefont {G.}~\bibnamefont {McCracken}}, \bibinfo
  {author} {\bibfnamefont {L.-M.}\ \bibnamefont {Duan}}, \bibinfo {author}
  {\bibfnamefont {P.}~\bibnamefont {Berman}}, \bibinfo {author} {\bibfnamefont
  {D.}~\bibnamefont {Steel}}, \bibinfo {author} {\bibfnamefont
  {A.}~\bibnamefont {Bracker}}, \bibinfo {author} {\bibfnamefont
  {D.}~\bibnamefont {Gammon}}, \ and\ \bibinfo {author} {\bibfnamefont
  {L.}~\bibnamefont {Sham}},\ }\href@noop {} {\bibfield  {journal} {\bibinfo
  {journal} {Physical review letters}\ }\textbf {\bibinfo {volume} {110}},\
  \bibinfo {pages} {167401} (\bibinfo {year} {2013})}\BibitemShut {NoStop}%
\bibitem [{\citenamefont {Reithmaier}\ \emph {et~al.}(2004)\citenamefont
  {Reithmaier}, \citenamefont {S{\^e}k}, \citenamefont {L{\"o}ffler},
  \citenamefont {Hofmann}, \citenamefont {Kuhn}, \citenamefont {Reitzenstein},
  \citenamefont {Keldysh}, \citenamefont {Kulakovskii}, \citenamefont
  {Reinecke},\ and\ \citenamefont {Forchel}}]{reithmaier2004}%
  \BibitemOpen
  \bibfield  {author} {\bibinfo {author} {\bibfnamefont {J.}~\bibnamefont
  {Reithmaier}}, \bibinfo {author} {\bibfnamefont {G.}~\bibnamefont {S{\^e}k}},
  \bibinfo {author} {\bibfnamefont {A.}~\bibnamefont {L{\"o}ffler}}, \bibinfo
  {author} {\bibfnamefont {C.}~\bibnamefont {Hofmann}}, \bibinfo {author}
  {\bibfnamefont {S.}~\bibnamefont {Kuhn}}, \bibinfo {author} {\bibfnamefont
  {S.}~\bibnamefont {Reitzenstein}}, \bibinfo {author} {\bibfnamefont
  {L.}~\bibnamefont {Keldysh}}, \bibinfo {author} {\bibfnamefont
  {V.}~\bibnamefont {Kulakovskii}}, \bibinfo {author} {\bibfnamefont
  {T.}~\bibnamefont {Reinecke}}, \ and\ \bibinfo {author} {\bibfnamefont
  {A.}~\bibnamefont {Forchel}},\ }\href@noop {} {\bibfield  {journal} {\bibinfo
   {journal} {Nature}\ }\textbf {\bibinfo {volume} {432}},\ \bibinfo {pages}
  {197} (\bibinfo {year} {2004})}\BibitemShut {NoStop}%
\bibitem [{\citenamefont {Faraon}\ \emph {et~al.}(2008)\citenamefont {Faraon},
  \citenamefont {Fushman}, \citenamefont {Englund}, \citenamefont {Stoltz},
  \citenamefont {Petroff},\ and\ \citenamefont
  {Vu{\v{c}}kovi{\'c}}}]{Faraon2008}%
  \BibitemOpen
  \bibfield  {author} {\bibinfo {author} {\bibfnamefont {A.}~\bibnamefont
  {Faraon}}, \bibinfo {author} {\bibfnamefont {I.}~\bibnamefont {Fushman}},
  \bibinfo {author} {\bibfnamefont {D.}~\bibnamefont {Englund}}, \bibinfo
  {author} {\bibfnamefont {N.}~\bibnamefont {Stoltz}}, \bibinfo {author}
  {\bibfnamefont {P.}~\bibnamefont {Petroff}}, \ and\ \bibinfo {author}
  {\bibfnamefont {J.}~\bibnamefont {Vu{\v{c}}kovi{\'c}}},\ }\href@noop {}
  {\bibfield  {journal} {\bibinfo  {journal} {Nature Physics}\ }\textbf
  {\bibinfo {volume} {4}},\ \bibinfo {pages} {859} (\bibinfo {year}
  {2008})}\BibitemShut {NoStop}%
\bibitem [{\citenamefont {Hohenester}\ \emph {et~al.}(2009)\citenamefont
  {Hohenester}, \citenamefont {Laucht}, \citenamefont {Kaniber}, \citenamefont
  {Hauke}, \citenamefont {Neumann}, \citenamefont {Mohtashami}, \citenamefont
  {Seliger}, \citenamefont {Bichler},\ and\ \citenamefont
  {Finley}}]{Laucht2009}%
  \BibitemOpen
  \bibfield  {author} {\bibinfo {author} {\bibfnamefont {U.}~\bibnamefont
  {Hohenester}}, \bibinfo {author} {\bibfnamefont {A.}~\bibnamefont {Laucht}},
  \bibinfo {author} {\bibfnamefont {M.}~\bibnamefont {Kaniber}}, \bibinfo
  {author} {\bibfnamefont {N.}~\bibnamefont {Hauke}}, \bibinfo {author}
  {\bibfnamefont {A.}~\bibnamefont {Neumann}}, \bibinfo {author} {\bibfnamefont
  {A.}~\bibnamefont {Mohtashami}}, \bibinfo {author} {\bibfnamefont
  {M.}~\bibnamefont {Seliger}}, \bibinfo {author} {\bibfnamefont
  {M.}~\bibnamefont {Bichler}}, \ and\ \bibinfo {author} {\bibfnamefont
  {J.~J.}\ \bibnamefont {Finley}},\ }\href {\doibase
  10.1103/PhysRevB.80.201311} {\bibfield  {journal} {\bibinfo  {journal} {Phys.
  Rev. B}\ }\textbf {\bibinfo {volume} {80}},\ \bibinfo {pages} {201311}
  (\bibinfo {year} {2009})}\BibitemShut {NoStop}%
\bibitem [{\citenamefont {Nakaoka}\ \emph {et~al.}(2006)\citenamefont
  {Nakaoka}, \citenamefont {Clark}, \citenamefont {Krenner}, \citenamefont
  {Sabathil}, \citenamefont {Bichler}, \citenamefont {Arakawa}, \citenamefont
  {Abstreiter},\ and\ \citenamefont {Finley}}]{Nakaoka2006}%
  \BibitemOpen
  \bibfield  {author} {\bibinfo {author} {\bibfnamefont {T.}~\bibnamefont
  {Nakaoka}}, \bibinfo {author} {\bibfnamefont {E.}~\bibnamefont {Clark}},
  \bibinfo {author} {\bibfnamefont {H.}~\bibnamefont {Krenner}}, \bibinfo
  {author} {\bibfnamefont {M.}~\bibnamefont {Sabathil}}, \bibinfo {author}
  {\bibfnamefont {M.}~\bibnamefont {Bichler}}, \bibinfo {author} {\bibfnamefont
  {Y.}~\bibnamefont {Arakawa}}, \bibinfo {author} {\bibfnamefont
  {G.}~\bibnamefont {Abstreiter}}, \ and\ \bibinfo {author} {\bibfnamefont
  {J.}~\bibnamefont {Finley}},\ }\href@noop {} {\bibfield  {journal} {\bibinfo
  {journal} {Physical Review B}\ }\textbf {\bibinfo {volume} {74}},\ \bibinfo
  {pages} {121305} (\bibinfo {year} {2006})}\BibitemShut {NoStop}%
\bibitem [{\citenamefont {M{\"u}ller}\ \emph
  {et~al.}(2012{\natexlab{a}})\citenamefont {M{\"u}ller}, \citenamefont
  {Bechtold}, \citenamefont {Ruppert}, \citenamefont {Zecherle}, \citenamefont
  {Reithmaier}, \citenamefont {Bichler}, \citenamefont {Krenner}, \citenamefont
  {Abstreiter}, \citenamefont {Holleitner}, \citenamefont {Villas-Boas} \emph
  {et~al.}}]{Muller2012phonon}%
  \BibitemOpen
  \bibfield  {author} {\bibinfo {author} {\bibfnamefont {K.}~\bibnamefont
  {M{\"u}ller}}, \bibinfo {author} {\bibfnamefont {A.}~\bibnamefont
  {Bechtold}}, \bibinfo {author} {\bibfnamefont {C.}~\bibnamefont {Ruppert}},
  \bibinfo {author} {\bibfnamefont {M.}~\bibnamefont {Zecherle}}, \bibinfo
  {author} {\bibfnamefont {G.}~\bibnamefont {Reithmaier}}, \bibinfo {author}
  {\bibfnamefont {M.}~\bibnamefont {Bichler}}, \bibinfo {author} {\bibfnamefont
  {H.}~\bibnamefont {Krenner}}, \bibinfo {author} {\bibfnamefont
  {G.}~\bibnamefont {Abstreiter}}, \bibinfo {author} {\bibfnamefont
  {A.}~\bibnamefont {Holleitner}}, \bibinfo {author} {\bibfnamefont
  {J.}~\bibnamefont {Villas-Boas}},  \emph {et~al.},\ }\href@noop {} {\bibfield
   {journal} {\bibinfo  {journal} {Physical review letters}\ }\textbf {\bibinfo
  {volume} {108}},\ \bibinfo {pages} {197402} (\bibinfo {year}
  {2012}{\natexlab{a}})}\BibitemShut {NoStop}%
\bibitem [{\citenamefont {F{\"o}rstner}\ \emph {et~al.}(2003)\citenamefont
  {F{\"o}rstner}, \citenamefont {Weber}, \citenamefont {Danckwerts},\ and\
  \citenamefont {Knorr}}]{Forstner2003}%
  \BibitemOpen
  \bibfield  {author} {\bibinfo {author} {\bibfnamefont {J.}~\bibnamefont
  {F{\"o}rstner}}, \bibinfo {author} {\bibfnamefont {C.}~\bibnamefont {Weber}},
  \bibinfo {author} {\bibfnamefont {J.}~\bibnamefont {Danckwerts}}, \ and\
  \bibinfo {author} {\bibfnamefont {A.}~\bibnamefont {Knorr}},\ }\href@noop {}
  {\bibfield  {journal} {\bibinfo  {journal} {Physical review letters}\
  }\textbf {\bibinfo {volume} {91}},\ \bibinfo {pages} {127401} (\bibinfo
  {year} {2003})}\BibitemShut {NoStop}%
\bibitem [{\citenamefont {Ramsay}\ \emph {et~al.}(2010)\citenamefont {Ramsay},
  \citenamefont {Gopal}, \citenamefont {Gauger}, \citenamefont {Nazir},
  \citenamefont {Lovett}, \citenamefont {Fox},\ and\ \citenamefont
  {Skolnick}}]{Ramsay2010}%
  \BibitemOpen
  \bibfield  {author} {\bibinfo {author} {\bibfnamefont {A.}~\bibnamefont
  {Ramsay}}, \bibinfo {author} {\bibfnamefont {A.~V.}\ \bibnamefont {Gopal}},
  \bibinfo {author} {\bibfnamefont {E.}~\bibnamefont {Gauger}}, \bibinfo
  {author} {\bibfnamefont {A.}~\bibnamefont {Nazir}}, \bibinfo {author}
  {\bibfnamefont {B.}~\bibnamefont {Lovett}}, \bibinfo {author} {\bibfnamefont
  {A.}~\bibnamefont {Fox}}, \ and\ \bibinfo {author} {\bibfnamefont
  {M.}~\bibnamefont {Skolnick}},\ }\href@noop {} {\bibfield  {journal}
  {\bibinfo  {journal} {Physical review letters}\ }\textbf {\bibinfo {volume}
  {104}},\ \bibinfo {pages} {017402} (\bibinfo {year} {2010})}\BibitemShut
  {NoStop}%
\bibitem [{\citenamefont {M{\"u}ller}\ \emph
  {et~al.}(2012{\natexlab{b}})\citenamefont {M{\"u}ller}, \citenamefont
  {Bechtold}, \citenamefont {Ruppert}, \citenamefont {Hautmann}, \citenamefont
  {Wildmann}, \citenamefont {Kaldewey}, \citenamefont {Bichler}, \citenamefont
  {Krenner}, \citenamefont {Abstreiter}, \citenamefont {Betz} \emph
  {et~al.}}]{muller2012high}%
  \BibitemOpen
  \bibfield  {author} {\bibinfo {author} {\bibfnamefont {K.}~\bibnamefont
  {M{\"u}ller}}, \bibinfo {author} {\bibfnamefont {A.}~\bibnamefont
  {Bechtold}}, \bibinfo {author} {\bibfnamefont {C.}~\bibnamefont {Ruppert}},
  \bibinfo {author} {\bibfnamefont {C.}~\bibnamefont {Hautmann}}, \bibinfo
  {author} {\bibfnamefont {J.}~\bibnamefont {Wildmann}}, \bibinfo {author}
  {\bibfnamefont {T.}~\bibnamefont {Kaldewey}}, \bibinfo {author}
  {\bibfnamefont {M.}~\bibnamefont {Bichler}}, \bibinfo {author} {\bibfnamefont
  {H.}~\bibnamefont {Krenner}}, \bibinfo {author} {\bibfnamefont
  {G.}~\bibnamefont {Abstreiter}}, \bibinfo {author} {\bibfnamefont
  {M.}~\bibnamefont {Betz}},  \emph {et~al.},\ }\href@noop {} {\bibfield
  {journal} {\bibinfo  {journal} {Physical Review B}\ }\textbf {\bibinfo
  {volume} {85}},\ \bibinfo {pages} {241306} (\bibinfo {year}
  {2012}{\natexlab{b}})}\BibitemShut {NoStop}%
\bibitem [{\citenamefont {Petta}\ \emph {et~al.}(2004)\citenamefont {Petta},
  \citenamefont {Johnson}, \citenamefont {Marcus}, \citenamefont {Hanson},\
  and\ \citenamefont {Gossard}}]{Petta2004}%
  \BibitemOpen
  \bibfield  {author} {\bibinfo {author} {\bibfnamefont {J.}~\bibnamefont
  {Petta}}, \bibinfo {author} {\bibfnamefont {A.}~\bibnamefont {Johnson}},
  \bibinfo {author} {\bibfnamefont {C.}~\bibnamefont {Marcus}}, \bibinfo
  {author} {\bibfnamefont {M.}~\bibnamefont {Hanson}}, \ and\ \bibinfo {author}
  {\bibfnamefont {A.}~\bibnamefont {Gossard}},\ }\href@noop {} {\bibfield
  {journal} {\bibinfo  {journal} {Physical review letters}\ }\textbf {\bibinfo
  {volume} {93}},\ \bibinfo {pages} {186802} (\bibinfo {year}
  {2004})}\BibitemShut {NoStop}%
\bibitem [{\citenamefont {Gl{\"a}ssl}\ \emph {et~al.}(2013)\citenamefont
  {Gl{\"a}ssl}, \citenamefont {Barth},\ and\ \citenamefont {Axt}}]{glassl2013}%
  \BibitemOpen
  \bibfield  {author} {\bibinfo {author} {\bibfnamefont {M.}~\bibnamefont
  {Gl{\"a}ssl}}, \bibinfo {author} {\bibfnamefont {A.}~\bibnamefont {Barth}}, \
  and\ \bibinfo {author} {\bibfnamefont {V.}~\bibnamefont {Axt}},\ }\href@noop
  {} {\bibfield  {journal} {\bibinfo  {journal} {Physical Review Letters}\
  }\textbf {\bibinfo {volume} {110}},\ \bibinfo {pages} {147401} (\bibinfo
  {year} {2013})}\BibitemShut {NoStop}%
\bibitem [{\citenamefont {Wei}\ \emph {et~al.}(2014)\citenamefont {Wei},
  \citenamefont {He}, \citenamefont {Chen}, \citenamefont {Hu}, \citenamefont
  {He}, \citenamefont {Wu}, \citenamefont {Schneider}, \citenamefont {Kamp},
  \citenamefont {H{\"o}fling}, \citenamefont {Lu} \emph {et~al.}}]{Wei2014}%
  \BibitemOpen
  \bibfield  {author} {\bibinfo {author} {\bibfnamefont {Y.-J.}\ \bibnamefont
  {Wei}}, \bibinfo {author} {\bibfnamefont {Y.-M.}\ \bibnamefont {He}},
  \bibinfo {author} {\bibfnamefont {M.-C.}\ \bibnamefont {Chen}}, \bibinfo
  {author} {\bibfnamefont {Y.-N.}\ \bibnamefont {Hu}}, \bibinfo {author}
  {\bibfnamefont {Y.}~\bibnamefont {He}}, \bibinfo {author} {\bibfnamefont
  {D.}~\bibnamefont {Wu}}, \bibinfo {author} {\bibfnamefont {C.}~\bibnamefont
  {Schneider}}, \bibinfo {author} {\bibfnamefont {M.}~\bibnamefont {Kamp}},
  \bibinfo {author} {\bibfnamefont {S.}~\bibnamefont {H{\"o}fling}}, \bibinfo
  {author} {\bibfnamefont {C.-Y.}\ \bibnamefont {Lu}},  \emph {et~al.},\
  }\href@noop {} {\bibfield  {journal} {\bibinfo  {journal} {arXiv preprint
  arXiv:1405.1991}\ } (\bibinfo {year} {2014})}\BibitemShut {NoStop}%
\bibitem [{\citenamefont {Mathew}\ \emph {et~al.}(2014)\citenamefont {Mathew},
  \citenamefont {Dilcher}, \citenamefont {Gamouras}, \citenamefont
  {Ramachandran}, \citenamefont {Yang}, \citenamefont {Freisem}, \citenamefont
  {Deppe},\ and\ \citenamefont {Hall}}]{Mathew2014}%
  \BibitemOpen
  \bibfield  {author} {\bibinfo {author} {\bibfnamefont {R.}~\bibnamefont
  {Mathew}}, \bibinfo {author} {\bibfnamefont {E.}~\bibnamefont {Dilcher}},
  \bibinfo {author} {\bibfnamefont {A.}~\bibnamefont {Gamouras}}, \bibinfo
  {author} {\bibfnamefont {A.}~\bibnamefont {Ramachandran}}, \bibinfo {author}
  {\bibfnamefont {H.~Y.~S.}\ \bibnamefont {Yang}}, \bibinfo {author}
  {\bibfnamefont {S.}~\bibnamefont {Freisem}}, \bibinfo {author} {\bibfnamefont
  {D.}~\bibnamefont {Deppe}}, \ and\ \bibinfo {author} {\bibfnamefont {K.~C.}\
  \bibnamefont {Hall}},\ }\href@noop {} {\bibfield  {journal} {\bibinfo
  {journal} {Physical Review B}\ }\textbf {\bibinfo {volume} {90}},\ \bibinfo
  {pages} {035316} (\bibinfo {year} {2014})}\BibitemShut {NoStop}%
\bibitem [{\citenamefont {Quilter}\ \emph {et~al.}(2014)\citenamefont
  {Quilter}, \citenamefont {Brash}, \citenamefont {Liu}, \citenamefont
  {Gl{\"a}ssl}, \citenamefont {Barth}, \citenamefont {Axt}, \citenamefont
  {Ramsay}, \citenamefont {Skolnick},\ and\ \citenamefont {Fox}}]{Quilter2014}%
  \BibitemOpen
  \bibfield  {author} {\bibinfo {author} {\bibfnamefont {J.}~\bibnamefont
  {Quilter}}, \bibinfo {author} {\bibfnamefont {A.}~\bibnamefont {Brash}},
  \bibinfo {author} {\bibfnamefont {F.}~\bibnamefont {Liu}}, \bibinfo {author}
  {\bibfnamefont {M.}~\bibnamefont {Gl{\"a}ssl}}, \bibinfo {author}
  {\bibfnamefont {A.}~\bibnamefont {Barth}}, \bibinfo {author} {\bibfnamefont
  {V.}~\bibnamefont {Axt}}, \bibinfo {author} {\bibfnamefont {A.}~\bibnamefont
  {Ramsay}}, \bibinfo {author} {\bibfnamefont {M.}~\bibnamefont {Skolnick}}, \
  and\ \bibinfo {author} {\bibfnamefont {A.}~\bibnamefont {Fox}},\ }\href@noop
  {} {\bibfield  {journal} {\bibinfo  {journal} {arXiv preprint
  arXiv:1409.0913}\ } (\bibinfo {year} {2014})}\BibitemShut {NoStop}%
\bibitem [{\citenamefont {Bounouar}\ \emph {et~al.}(2014)\citenamefont
  {Bounouar}, \citenamefont {M{\"u}ller}, \citenamefont {Barth}, \citenamefont
  {Gl{\"a}ssl}, \citenamefont {Axt},\ and\ \citenamefont
  {Michler}}]{Bounouar2014}%
  \BibitemOpen
  \bibfield  {author} {\bibinfo {author} {\bibfnamefont {S.}~\bibnamefont
  {Bounouar}}, \bibinfo {author} {\bibfnamefont {M.}~\bibnamefont
  {M{\"u}ller}}, \bibinfo {author} {\bibfnamefont {A.}~\bibnamefont {Barth}},
  \bibinfo {author} {\bibfnamefont {M.}~\bibnamefont {Gl{\"a}ssl}}, \bibinfo
  {author} {\bibfnamefont {V.}~\bibnamefont {Axt}}, \ and\ \bibinfo {author}
  {\bibfnamefont {P.}~\bibnamefont {Michler}},\ }\href@noop {} {\bibfield
  {journal} {\bibinfo  {journal} {arXiv preprint arXiv:1408.7027}\ } (\bibinfo
  {year} {2014})}\BibitemShut {NoStop}%
\bibitem [{\citenamefont {Warburton}\ \emph {et~al.}(2000)\citenamefont
  {Warburton}, \citenamefont {Sch{\"a}flein}, \citenamefont {Haft},
  \citenamefont {Bickel}, \citenamefont {Lorke}, \citenamefont {Karrai},
  \citenamefont {Garcia}, \citenamefont {Schoenfeld},\ and\ \citenamefont
  {Petroff}}]{warburton2000}%
  \BibitemOpen
  \bibfield  {author} {\bibinfo {author} {\bibfnamefont {R.~J.}\ \bibnamefont
  {Warburton}}, \bibinfo {author} {\bibfnamefont {C.}~\bibnamefont
  {Sch{\"a}flein}}, \bibinfo {author} {\bibfnamefont {D.}~\bibnamefont {Haft}},
  \bibinfo {author} {\bibfnamefont {F.}~\bibnamefont {Bickel}}, \bibinfo
  {author} {\bibfnamefont {A.}~\bibnamefont {Lorke}}, \bibinfo {author}
  {\bibfnamefont {K.}~\bibnamefont {Karrai}}, \bibinfo {author} {\bibfnamefont
  {J.~M.}\ \bibnamefont {Garcia}}, \bibinfo {author} {\bibfnamefont
  {W.}~\bibnamefont {Schoenfeld}}, \ and\ \bibinfo {author} {\bibfnamefont
  {P.~M.}\ \bibnamefont {Petroff}},\ }\href@noop {} {\bibfield  {journal}
  {\bibinfo  {journal} {Nature}\ }\textbf {\bibinfo {volume} {405}},\ \bibinfo
  {pages} {926} (\bibinfo {year} {2000})}\BibitemShut {NoStop}%
\bibitem [{\citenamefont {Kuhlmann}\ \emph {et~al.}(2013)\citenamefont
  {Kuhlmann}, \citenamefont {Houel}, \citenamefont {Brunner}, \citenamefont
  {Ludwig}, \citenamefont {Reuter}, \citenamefont {Wieck},\ and\ \citenamefont
  {Warburton}}]{Kuhlmann2013}%
  \BibitemOpen
  \bibfield  {author} {\bibinfo {author} {\bibfnamefont {A.~V.}\ \bibnamefont
  {Kuhlmann}}, \bibinfo {author} {\bibfnamefont {J.}~\bibnamefont {Houel}},
  \bibinfo {author} {\bibfnamefont {D.}~\bibnamefont {Brunner}}, \bibinfo
  {author} {\bibfnamefont {A.}~\bibnamefont {Ludwig}}, \bibinfo {author}
  {\bibfnamefont {D.}~\bibnamefont {Reuter}}, \bibinfo {author} {\bibfnamefont
  {A.~D.}\ \bibnamefont {Wieck}}, \ and\ \bibinfo {author} {\bibfnamefont
  {R.~J.}\ \bibnamefont {Warburton}},\ }\href@noop {} {\bibfield  {journal}
  {\bibinfo  {journal} {Review of Scientific Instruments}\ }\textbf {\bibinfo
  {volume} {84}},\ \bibinfo {pages} {073905} (\bibinfo {year}
  {2013})}\BibitemShut {NoStop}%
\bibitem [{\citenamefont {He}\ \emph {et~al.}(2013)\citenamefont {He},
  \citenamefont {He}, \citenamefont {Wei}, \citenamefont {Wu}, \citenamefont
  {Atat{\"u}re}, \citenamefont {Schneider}, \citenamefont {H{\"o}fling},
  \citenamefont {Kamp}, \citenamefont {Lu},\ and\ \citenamefont
  {Pan}}]{He2013}%
  \BibitemOpen
  \bibfield  {author} {\bibinfo {author} {\bibfnamefont {Y.-M.}\ \bibnamefont
  {He}}, \bibinfo {author} {\bibfnamefont {Y.}~\bibnamefont {He}}, \bibinfo
  {author} {\bibfnamefont {Y.-J.}\ \bibnamefont {Wei}}, \bibinfo {author}
  {\bibfnamefont {D.}~\bibnamefont {Wu}}, \bibinfo {author} {\bibfnamefont
  {M.}~\bibnamefont {Atat{\"u}re}}, \bibinfo {author} {\bibfnamefont
  {C.}~\bibnamefont {Schneider}}, \bibinfo {author} {\bibfnamefont
  {S.}~\bibnamefont {H{\"o}fling}}, \bibinfo {author} {\bibfnamefont
  {M.}~\bibnamefont {Kamp}}, \bibinfo {author} {\bibfnamefont {C.-Y.}\
  \bibnamefont {Lu}}, \ and\ \bibinfo {author} {\bibfnamefont {J.-W.}\
  \bibnamefont {Pan}},\ }\href@noop {} {\bibfield  {journal} {\bibinfo
  {journal} {Nature nanotechnology}\ }\textbf {\bibinfo {volume} {8}},\
  \bibinfo {pages} {213} (\bibinfo {year} {2013})}\BibitemShut {NoStop}%
\bibitem [{\citenamefont {Zecherle}\ \emph {et~al.}(2010)\citenamefont
  {Zecherle}, \citenamefont {Ruppert}, \citenamefont {Clark}, \citenamefont
  {Abstreiter}, \citenamefont {Finley},\ and\ \citenamefont
  {Betz}}]{Zecherle2010}%
  \BibitemOpen
  \bibfield  {author} {\bibinfo {author} {\bibfnamefont {M.}~\bibnamefont
  {Zecherle}}, \bibinfo {author} {\bibfnamefont {C.}~\bibnamefont {Ruppert}},
  \bibinfo {author} {\bibfnamefont {E.}~\bibnamefont {Clark}}, \bibinfo
  {author} {\bibfnamefont {G.}~\bibnamefont {Abstreiter}}, \bibinfo {author}
  {\bibfnamefont {J.}~\bibnamefont {Finley}}, \ and\ \bibinfo {author}
  {\bibfnamefont {M.}~\bibnamefont {Betz}},\ }\href@noop {} {\bibfield
  {journal} {\bibinfo  {journal} {Physical Review B}\ }\textbf {\bibinfo
  {volume} {82}},\ \bibinfo {pages} {125314} (\bibinfo {year}
  {2010})}\BibitemShut {NoStop}%
\bibitem [{\citenamefont {Vagov}\ \emph {et~al.}(2011)\citenamefont {Vagov},
  \citenamefont {Croitoru}, \citenamefont {Gl{\"a}ssl}, \citenamefont {Axt},\
  and\ \citenamefont {Kuhn}}]{Vagov2011}%
  \BibitemOpen
  \bibfield  {author} {\bibinfo {author} {\bibfnamefont {A.}~\bibnamefont
  {Vagov}}, \bibinfo {author} {\bibfnamefont {M.}~\bibnamefont {Croitoru}},
  \bibinfo {author} {\bibfnamefont {M.}~\bibnamefont {Gl{\"a}ssl}}, \bibinfo
  {author} {\bibfnamefont {V.}~\bibnamefont {Axt}}, \ and\ \bibinfo {author}
  {\bibfnamefont {T.}~\bibnamefont {Kuhn}},\ }\href@noop {} {\bibfield
  {journal} {\bibinfo  {journal} {Physical Review B}\ }\textbf {\bibinfo
  {volume} {83}},\ \bibinfo {pages} {094303} (\bibinfo {year}
  {2011})}\BibitemShut {NoStop}%
\bibitem [{\citenamefont {Johansson}\ \emph {et~al.}(2012)\citenamefont
  {Johansson}, \citenamefont {Nation},\ and\ \citenamefont {Nori}}]{qutip}%
  \BibitemOpen
  \bibfield  {author} {\bibinfo {author} {\bibfnamefont {J.}~\bibnamefont
  {Johansson}}, \bibinfo {author} {\bibfnamefont {P.}~\bibnamefont {Nation}}, \
  and\ \bibinfo {author} {\bibfnamefont {F.}~\bibnamefont {Nori}},\ }\href@noop
  {} {\bibfield  {journal} {\bibinfo  {journal} {Computer Physics
  Communications}\ }\textbf {\bibinfo {volume} {183}},\ \bibinfo {pages} {1760}
  (\bibinfo {year} {2012})}\BibitemShut {NoStop}%
\bibitem [{\citenamefont {Cohen-Tannoudji}\ \emph {et~al.}(1992)\citenamefont
  {Cohen-Tannoudji}, \citenamefont {Dupont-Roc},\ and\ \citenamefont
  {Grynberg}}]{cohen1992}%
  \BibitemOpen
  \bibfield  {author} {\bibinfo {author} {\bibfnamefont {C.}~\bibnamefont
  {Cohen-Tannoudji}}, \bibinfo {author} {\bibfnamefont {J.}~\bibnamefont
  {Dupont-Roc}}, \ and\ \bibinfo {author} {\bibfnamefont {G.}~\bibnamefont
  {Grynberg}},\ }\href@noop {} {\emph {\bibinfo {title} {Atom-photon
  interactions: basic processes and applications}}}\ (\bibinfo  {publisher}
  {Wiley Online Library},\ \bibinfo {year} {1992})\BibitemShut {NoStop}%
\bibitem [{\citenamefont {Gl{\"a}ssl}\ \emph {et~al.}(2011)\citenamefont
  {Gl{\"a}ssl}, \citenamefont {Vagov}, \citenamefont {L{\"u}ker}, \citenamefont
  {Reiter}, \citenamefont {Croitoru}, \citenamefont {Machnikowski},
  \citenamefont {Axt},\ and\ \citenamefont {Kuhn}}]{glassl2011}%
  \BibitemOpen
  \bibfield  {author} {\bibinfo {author} {\bibfnamefont {M.}~\bibnamefont
  {Gl{\"a}ssl}}, \bibinfo {author} {\bibfnamefont {A.}~\bibnamefont {Vagov}},
  \bibinfo {author} {\bibfnamefont {S.}~\bibnamefont {L{\"u}ker}}, \bibinfo
  {author} {\bibfnamefont {D.}~\bibnamefont {Reiter}}, \bibinfo {author}
  {\bibfnamefont {M.}~\bibnamefont {Croitoru}}, \bibinfo {author}
  {\bibfnamefont {P.}~\bibnamefont {Machnikowski}}, \bibinfo {author}
  {\bibfnamefont {V.}~\bibnamefont {Axt}}, \ and\ \bibinfo {author}
  {\bibfnamefont {T.}~\bibnamefont {Kuhn}},\ }\href@noop {} {\bibfield
  {journal} {\bibinfo  {journal} {Physical Review B}\ }\textbf {\bibinfo
  {volume} {84}},\ \bibinfo {pages} {195311} (\bibinfo {year}
  {2011})}\BibitemShut {NoStop}%
\bibitem [{\citenamefont {Stufler}\ \emph {et~al.}(2006)\citenamefont
  {Stufler}, \citenamefont {Machnikowski}, \citenamefont {Ester}, \citenamefont
  {Bichler}, \citenamefont {Axt}, \citenamefont {Kuhn},\ and\ \citenamefont
  {Zrenner}}]{Stufler2006}%
  \BibitemOpen
  \bibfield  {author} {\bibinfo {author} {\bibfnamefont {S.}~\bibnamefont
  {Stufler}}, \bibinfo {author} {\bibfnamefont {P.}~\bibnamefont
  {Machnikowski}}, \bibinfo {author} {\bibfnamefont {P.}~\bibnamefont {Ester}},
  \bibinfo {author} {\bibfnamefont {M.}~\bibnamefont {Bichler}}, \bibinfo
  {author} {\bibfnamefont {V.}~\bibnamefont {Axt}}, \bibinfo {author}
  {\bibfnamefont {T.}~\bibnamefont {Kuhn}}, \ and\ \bibinfo {author}
  {\bibfnamefont {A.}~\bibnamefont {Zrenner}},\ }\href@noop {} {\bibfield
  {journal} {\bibinfo  {journal} {Physical Review B}\ }\textbf {\bibinfo
  {volume} {73}},\ \bibinfo {pages} {125304} (\bibinfo {year}
  {2006})}\BibitemShut {NoStop}%
\end{thebibliography}%
\end{document}